\documentclass[twocolumn]{article}

\usepackage{bm,subfigure,graphicx,cite}

\title{Meta-stable nematic pre-ordering in smectic liquid crystalline phase transitions}
\author{Nasser Mohieddin Abukhdeir and Alejandro D. Rey\\
nasser.abukhdeir@mcgill.ca, alejandro.rey@mcgill.ca\\
Department of Chemical Engineering\\
McGill University, Montreal, Quebec, Canada}
\date{\today}

\begin{document}

\maketitle

% ABSTRACT (not included in Communication directly)
%
% Modeling and simulation studies using a high-order Landau-de Gennes type model for the direct isotropic/smectic-A liquid crystal transition are presented showing the existence of meta-stable nematic pre-ordering under certain conditions, in agreement with experimental observations (Tokita, M. et al. Macromolecules 2006, 39, 2021–2023)

The role of liquid crystalline mesophases in polymer crystallization has seen increased attention following the observations of meta-stable liquid crystalline pre-ordering in the crystallization of isotactic polypropylene and other flexible polymers \cite{Imai1993,Imai1994,Matsuba1999,Asano1999,Ran2002,Li2004b,Soccio2007}.  In addition to these recent observations, both in Nature and industry liquid crystal mesophases are used to create polymeric materials with desirable properties such as spider silk \cite{Vollrath2001} and Kevlar \cite{Dobb1977}. A better understanding of the theoretical fundamentals behind formation of the formation kinetics of polymer mesophases and their subsequent effect on crystallization would enable many new materials and processing techniques to be developed.  

A recent study by Tokita et al \cite{Tokita2006} on a polymer liquid crystal that exhibits a direct isotropic/smectic transition has experimentally determined the existence of meta-stable nematic orientational ordering preceding the formation of translational smectic ordering.  This was accomplished through the study of a polymer liquid crystalline material BB-3(1-ME) that exhibits extremely slow liquid crystalline transition dynamics, also developed by Tokita et al \cite{Tokita2004}.  The phase-ordering dynamics of this material occur at time scales accessible via conventional methods of polarized light scattering and synchrotron wide-angle X-ray diffraction analysis techniques \cite{Tokita2004,Tokita2006}. Two key conclusions were determined from this work.  First, at high-quench rates/super-cooling meta-stable nematic (orientational) ordering was observed preceding full smectic (orientational and translational) order. Second, the occurrence of nematic pre-ordering (high-supercooling) resulted in morphological changes of growing liquid crystalline domains compared to solely smectic growth (low super-cooling).  Specifically, samples cooled at rates high enough to exhibit nematic pre-ordering formed well-oriented or ``neat'' tactoidal smectic domains while samples cooled at lower rates, where only smectic ordering was observed, formed radially oriented or textured spherulitic domains. It is noted that tactoids are ubiquitous transient non-spherical drops frequently observed in phase transformations involving mesophases \cite{Dierking2003,Prinsen2004}. 

The objective  of this communication is to show that the essential features of the experimental observations of Tokita et al \cite{Tokita2006} described above emerge from mesophase phase transformation phenomena as described by well-established phenomenological Landau-de Gennes type models.

The existence of meta-stable pre-ordering in phase-ordering growth, where the order parameter is non-conserved (such as orientation), was first demonstrated theoretically by Bechhoefer et al \cite{Bechhoefer1991}, and later for dual non-conserved order parameter systems \cite{Tuckerman1992}.  In this past work, Landau-Ginzburg phase transition models composed of scalar order parameters were shown to exhibition a front-splitting instability under strong super-cooling and also predicted this to occur when large differences in transition time-scales exist.  

\begin{figure} 
\centering
\includegraphics[width=0.9\linewidth]{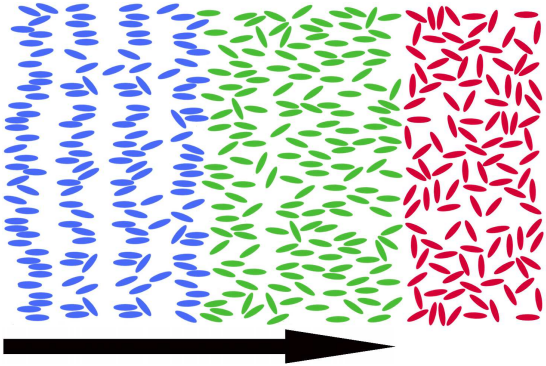}

\caption{A schematic of a growing liquid crystalline front summarizing the phenomena of interest: nematic (orientational) and smectic-A (lamellar) liquid crystalline ordering and interfacial splitting.  The orientationally/translationally-order smectic-A mesophase is on the left (blue), orientationally-ordered nematic mesophase is in the center (green), and the isotropic liquid (no orientational or translational order) is on the right (red).}
\label{fig:lcorder}
\end{figure}

The scalar models used by Bechhoefer et al \cite{Bechhoefer1991,Tuckerman1992} capture the basic pre-ordering phenomena but are not suitable for comparison to experimental observations.  This is especially true for liquid crystalline materials where phase-ordering introduces additional physics due to inherent anisotropy (see Figure \ref{fig:lcorder}).  Smectic liquid crystals include dual phase-ordering phenomena where orientational order is required for translational order to exist.  Thus with such a system three general effects influence the possible observation of meta-stable pre-ordering: (1) the driving potential (as a function of temperature, concentration, and pressure), (2) energetic coupling between the two types of order, and (3) phase transition kinetics (difference in time scales governed by viscosity ratio).

In this communication we demonstrate that the use of a high-order model and numerical simulation is able bridge the gap between the experimental observations of Tokita et al \cite{Tokita2006} and the fundamental observations of Bechhoefer et al \cite{Bechhoefer1991,Tuckerman1992}. As will be shown, this approach both captures meta-stable nematic pre-ordering in addition to actual multi-scale dynamic texturing processes.  A high-order Landau-de Gennes type phenomenological model \cite{deGennes1995,Mukherjee2001} for the direct isotropic/smectic-A liquid crystalline transition is used with material parameters based on the 12CB (dodecyl-cyanobiphenyl) liquid crystal.  This model does not take into account microstructure that is present in polymer liquid crystals, but is an adequate first-approximation for the system studied by Tokita et al \cite{Tokita2006}.  The variation of the time scales of nematic (orientational) and smectic-A (translational) ordering is easily achieved with this model and ideal quenches are assumed. The effects of changing the energetic coupling between the orientational and translation ordering of the smectic-A mesophase, in addition to convection and heat of transition, are left to future work in that determination of a suitable set of phenomenological constants and model extensions are non-trivial tasks \cite{Abukhdeir2008c}.

Theoretical characterization of orientational and translational order is accomplished using order parameters that adequately capture the physics involved.   Partial orientational order of the nematic phase is characterized using a symmetric traceless tensor \cite{deGennes1995}:\begin{equation} \label{eqn:nem_order_param}
\bm{Q} = S \left(\bm{nn} - \frac{1}{3} \bm{I}\right) + \frac{1}{3} P \left( \bm{mm} - \bm{ll}\right)
\end{equation}                   
where $\mathbf{n}/\mathbf{m}/\mathbf{l}$ are the eigenvectors of $\bm{Q}$, which characterize the average molecular orientational axes, and $S/P$ are scalars which characterize the extent to which the molecules conform to the average orientational axes \cite{Rey2002,Yan2002,Rey2007}.  Uniaxial order is characterized by $S$ and $\bm{n}$, which correspond to the maximum eigenvalue (and its corresponding eigenvector) of $\bm{Q}$, $S= \frac{3}{2} \mu_n$.  Biaxial order is characterized by $P$ and $\bm{m}/\bm{l}$, which correspond to the lesser eigenvalues and eigenvectors, $P = -\frac{3}{2}\left(\mu_m - \mu_l\right)$.

The one-dimensional translational order of the smectic-A mesophase in addition to the orientational order found in nematics is characterized through the use of primary (orientational) and secondary (translational) order parameters together \cite{Toledano1987}.  A complex order parameter can be used to characterize translational order \cite{deGennes1995}:
\begin{equation} \label{eqsmec_order_param}
\Psi = \psi e^{i \phi}
\end{equation}
where $\phi$ is the phase and $\psi$ is the scalar amplitude of the density modulation.  The density wave vector, which describes the average orientation of the smectic-A density modulation, is defined as $\mathbf{a} = \nabla \phi / |{\nabla \phi}|$.  The smectic scalar order parameter $\psi$ characterizes the magnitude of the density modulation and is used in a dimensionless form in this work.  In the smectic-A mesophase the preferred orientation of the wave vector is parallel to the average molecular orientational axis, $\mathbf{n}$.

A Landau-de Gennes type model for the first order isotropic/smectic-A phase transition is used that was initially presented by Mukherjee, Pleiner, and Brand \cite{deGennes1995,Mukherjee2001} and later extended by adding nematic elastic terms \cite{Brand2001,Mukherjee2002a,Biscari2007}:
\begin{eqnarray} \label{eq:free_energy_heterogeneous}
f - f_0 &=&\frac{1}{2} a \left(\bm{Q} : \bm{Q}\right) - \frac{1}{3} b \left(\bm{Q}\cdot\bm{Q}\right) : \bm{Q} + \frac{1}{4} c \left(\bm{Q} : \bm{Q}\right)^2  \nonumber\\
&& + \frac{1}{2} \alpha \left|\Psi\right|^2 + \frac{1}{4} \beta \left|\Psi\right|^4 \nonumber\\
&&- \frac{1}{2} \delta \psi^2 \left(\bm{Q} : \bm{Q}\right) - \frac{1}{2} e \bm{Q}:\left(\bm{\nabla} \Psi\right)\left(\bm{\nabla} \Psi^*\right) \nonumber\\
&& + \frac{1}{2} l_1 (\bm{\nabla} \bm{Q} \vdots \bm{\nabla} \bm{Q} ) + \frac{1}{2} l_1 (\bm{\nabla} \cdot \bm{Q} \cdot \bm{\nabla} \cdot \bm{Q} ) \nonumber\\
&& + \frac{1}{2} l_3 \bm{Q}:\left( \nabla \bm{Q} : \nabla \bm{Q} \right) \nonumber\\ 
&&+ \frac{1}{2} b_1 \left|\bm{\nabla} \Psi\right|^2 + \frac{1}{4} b_2 \left|\nabla^2 \Psi\right|^2
\end{eqnarray}
\begin{eqnarray} \label{eq:free_energy_heterogenous_coeffs}
a & = & a_0 (T - T_{NI}) \nonumber \\
\alpha & = & \alpha_0 (T - T_{AI})\nonumber 
\end{eqnarray}
where $f$ is the free energy density, $f_0$ is the free energy density of the isotropic phase, terms 1-5 are the bulk contributions to the free energy, terms 6-7 are couplings of nematic and smectic order; both the bulk order and coupling of the nematic director and smectic density-wave vector, respectively.  Terms 8-10/11-12 are the nematic/smectic elastic contributions to the free energy.  $T$ is temperature, $T_{NI}$/$T_{AI}$ are the hypothetical second order transition temperatures for isotropic/nematic and isotropic/smectic-A mesophase transitions (refer to \cite{Coles1979a} for more detail), and the remaining constants are phenomenological parameters.

The Landau-Ginzburg time-dependent formulation \cite{Barbero2000} is used to capture the dynamics of the phase transition. The general form of the time-dependent formulation is as follows \cite{Barbero2000}:
\begin{eqnarray} \label{eq:landau_ginz}
\left(\begin{array}{c}
 \frac{\partial \bm{Q}}{\partial t}
\\ \frac{\partial A}{\partial t}
\\ \frac{\partial B}{\partial t} 
\end{array}\right)
&=& 
\left(\begin{array}{c c c} 
\frac{1}{\mu_n} & 0 & 0\\ 
0 & \frac{1}{\mu_S} & 0\\ 
0 & 0 & \frac{1}{\mu_S}\end{array} \right)
\left(\begin{array}{c} -\frac{\delta F}{\delta \bm{Q}}\\ 
-\frac{\delta F}{\delta A}\\ 
-\frac{\delta F}{\delta B} \end{array}\right)\\
F &=& \int_V f dV
\end{eqnarray}
where $\mu_r$/$\mu_s$ is the rotational/smectic viscosity, and $V$ the volume. A higher order functional derivative must be used due to the second-derivative term in the free energy equation (\ref{eq:free_energy_heterogeneous}):
\begin{equation} \label{eqpdes}
\frac{\delta F}{\delta \theta} = \frac{\partial f}{\partial \theta} - \frac{\partial}{\partial x_i}\left(\frac{\partial f}{\partial \frac{\partial \theta}{\partial x_i}} \right) + \frac{\partial}{\partial x_i}\frac{\partial}{\partial x_j}\left(\frac{\partial f}{\partial \frac{\partial^2 \theta}{\partial x_i \partial x_j}} \right)
\end{equation}
where $\theta$ corresponds to the order parameter.

\begin{figure*}
\centering
\subfigure[]{\includegraphics[width=0.45\linewidth]{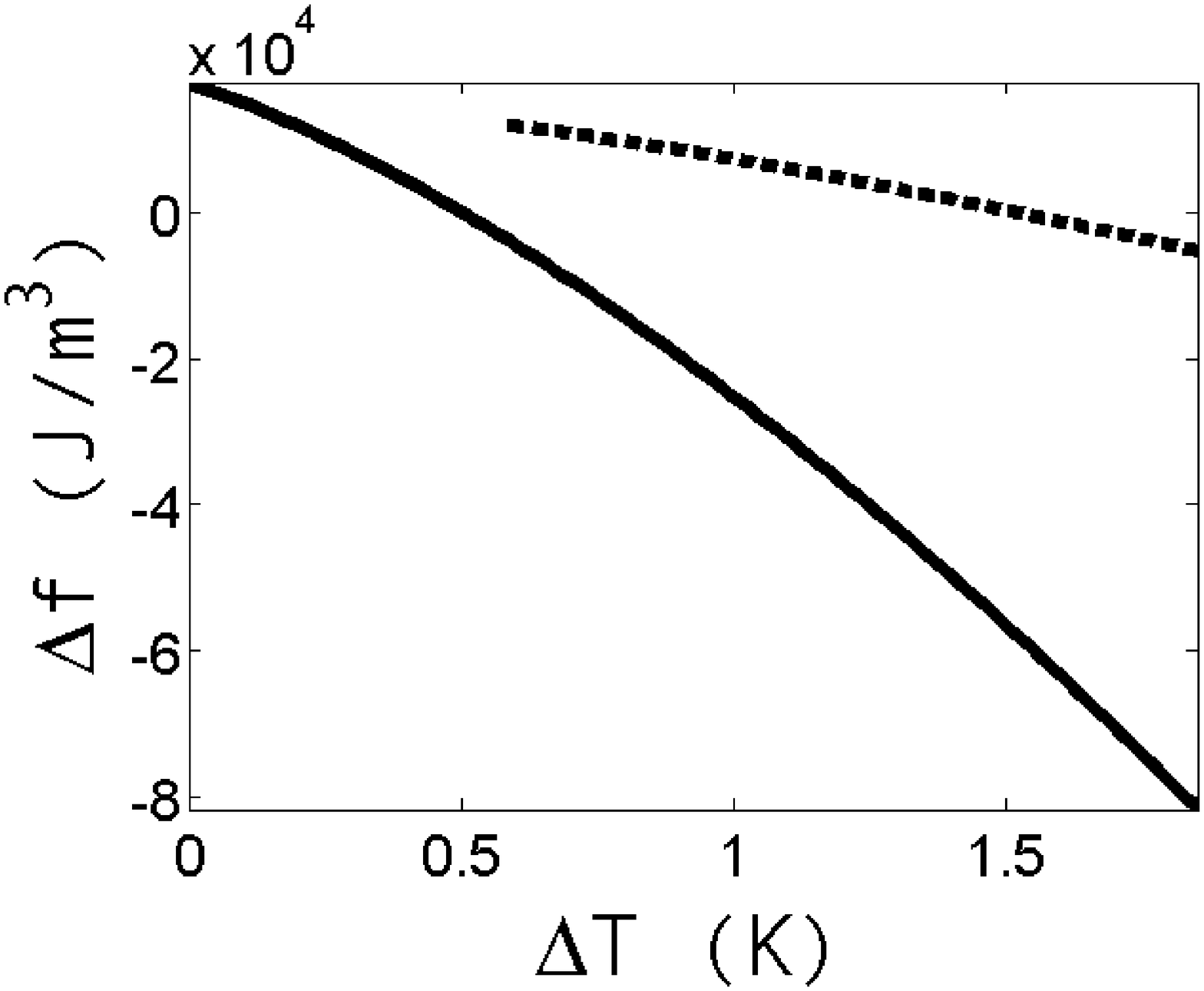}}
\subfigure[]{\includegraphics[width=0.45\linewidth]{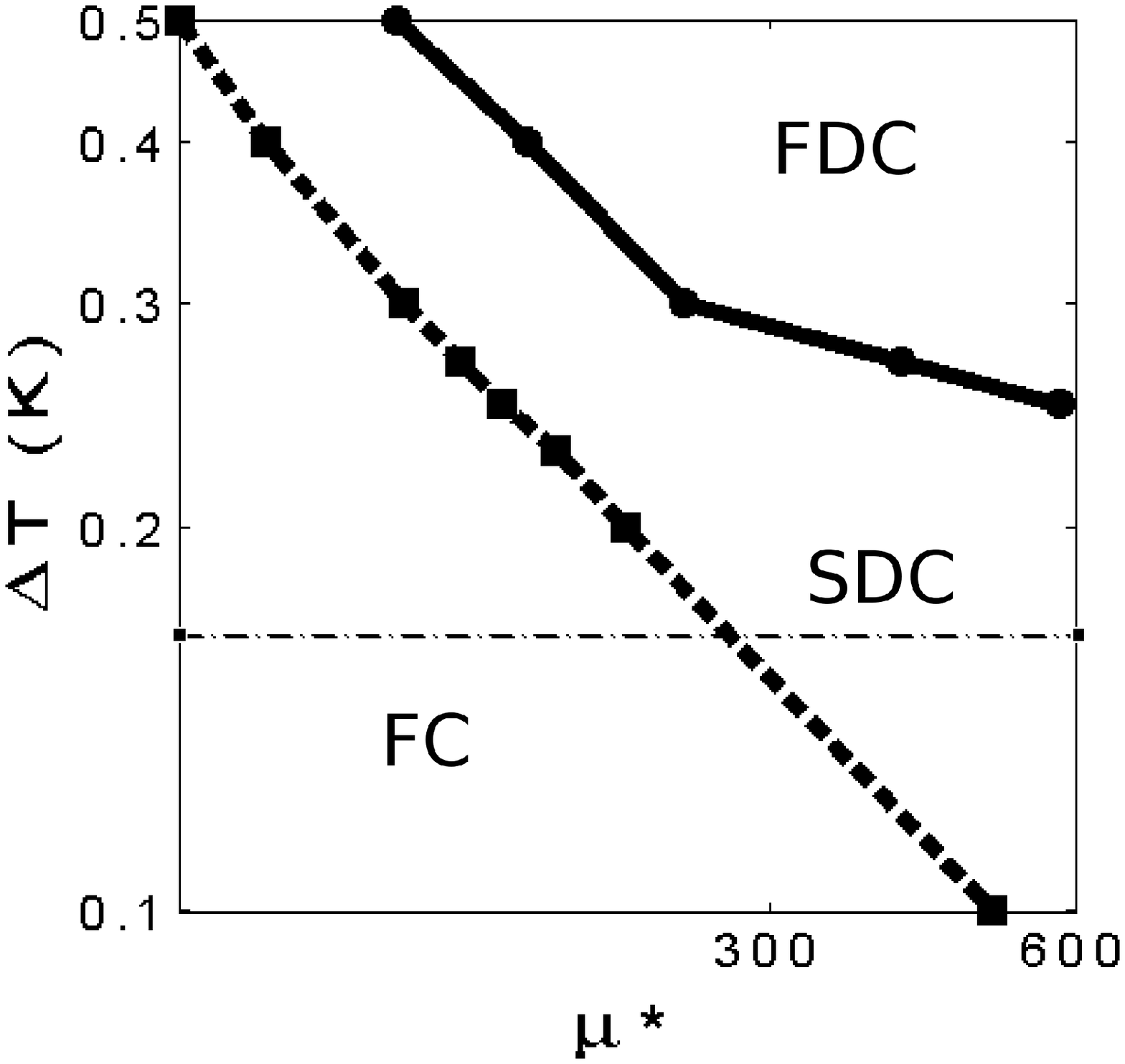}}
\caption{a) Plot of free energy density versus quench depth $\Delta = T_{AI} - T$ for neat smectic-A order (solid line) and degenerate nematic order (dotted line) b) Log-log plot of quench depth versus viscosity ratio $\mu^*= \frac{\mu_s}{\mu_n}$ (ratio of timescale of orientational and translation order) of the numerical determined SDC (squares/dotted line), FDC (circles/solid line) boundaries, and temperature at which the degenerate nematic phase becomes unstable (dot-dashed line); see text. The material parameters and phenomenological coefficients, based upon 12CB \cite{Abukhdeir2008c}, are $T_{NI}=322.85K$, $T_{AI}=330.5K$, $a_0= 2\times10^5\frac{J}{m^3 K}$, $b=2.823\times10^7\frac{J}{m^3}$, $c=1.972\times10^7\frac{J}{m^3}$, $\alpha_0=1.903\times10^6\frac{J}{m^3 K}$, $\beta=3.956\times10^8\frac{J}{m^3}$, $\delta=9.792\times10^6 \frac{J}{m^3}$, $e=1.938\times10^{-11}pN$, $l_1=1\times10^{-12}pN$, $b_1=1\times 10^{-12}pN$, $b_2=3.334\times10^{-30}Jm$, $\mu_N = 8.4\times10^{-2}\frac{N \times s}{m^2}$.}
\label{fig:plot}
\end{figure*}

As shown below, in this communication we capture the essential experimental results \cite{Tokita2006} by varying the quench depth  and  the translational/rotational viscosity ratio. This choice follows from the fact that the phase transformation velocity is given by ratio of the temperature-controlled free energy driving force and the viscosity associated with mesophase ordering \cite{Wincure2006}. This is first studied by treating the kinetic phase transformation process as a one-dimensional propagation transition front(s), where the unstable isotropic phase is replaced by the stable smectic-A phase. Multiple one-dimensional simulations were conducted for different viscosity ratio/quench depth values and the smectic-A/meta-stable nematic front behavior characterized. These results were then used to determine a dynamic phase diagram in order to guide higher-dimensional simulations. Following this, a pair of two-dimensional simulations were performed in regimes where meta-stable nematic ordering is (1) not present and (2) present, guided by the previously determined dynamic phase diagram. All computational results were achieved by the simulation of eqn \ref{eq:landau_ginz}, using the same computational methods defined in ref. \cite{Abukhdeir2008c}. These two stages of the presented work correspond directly to the two key conclusions of the experimental work of Tokita et al \cite{Tokita2006}.

Previous work for phenomenological parameter determination (parameters in eqn \ref{eq:free_energy_heterogeneous}) and phase diagram computation has been completed for this system \cite{Abukhdeir2008c} which shows that a definite free energy minimum exists corresponding to smectic-A order.  In addition, a degenerate nematic-order state exists below the theoretical second order transition temperature $T_{AI}$ for this system. Figure \ref{fig:plot}a shows the free energy density (\ref{eq:free_energy_heterogeneous}) versus quench depth ($\Delta T = T_{AI}-T$) for both neat smectic-A and degenerate nematic order as predicted by the model. As mentioned previously, one-dimensional front growth was studied by varying the difference in time scales for the formation of nematic and smectic-A order, $\mu^* = \frac{\mu_S}{\mu_N}$, and the magnitude of the driving variable, temperature quench depth $\Delta T$. Figure \ref{fig:plot}b shows the computed dynamic growth phase diagram where three distinct regimes can be identified:
\begin{enumerate}
\item Fully Coupled (FC, lower left region), in this regime nematic and smectic-A interfaces cannot be distinguished and corresponds to the direct transition from an isotropic to smectic-A phase. In other words, the nematic pre-ordering layer is not observed.
\item Static Decoupled (SDC, middle region), in this regime meta-stable nematic pre-ordering (see Figure \ref{fig:plot}a) is observed to proceed from the initial smectic-A front but maintains a constant distance/halo and does not fully decouple. In this case, a constant thickness nematic pre-ordering layer separates the unstable receding isotropic phase from the stable advancing smectic A phase. 
\item Fully Decoupled (FDC, upper right region), in this regime meta-stable nematic pre-ordering (see Figure \ref{fig:plot}a) is observed to proceed from the initial smectic-A front at a velocity greater than the preceding smectic-A front velocity. In this case, the pre-ordering nematic layer replaces the isotropic phase and becomes the matrix in which the smectic A phase grows.
\end{enumerate}

Next we identify the main mechanisms behind the results of Figure \ref{fig:plot}a. The effect of the quench depth $\Delta T$ on the kinetics of the phase-ordering transition is revealed by the work of Bechhoefer et al \cite{Bechhoefer1991,Tuckerman1992}. Whereas previous work on a dual non-conserved order parameter system focused on quench depth as the driving variable to induce the splitting instability \cite{Tuckerman1992}, the effect of differences in dynamic timescales of orientational and positional ordering can induce the same phenomenon.  In the case of low molecular mass liquid crystals, as this model is most applicable, experimental observations point towards the FC regime.  In the case of liquid crystalline polymers, where the time scale for local diffusion increases substantially compared to rotation (rigid/semi-rigid and side-chain liquid crystalline polymers), the contribution of the difference in timescales is predicted to exhibit this behavior at much lower quench depths, observed experimentally \cite{Tokita2006} and in agreement with the presented results (Figure \ref{fig:plot})b. 

\begin{figure*}
\centering 
\subfigure[]{\includegraphics[width=0.3\linewidth]{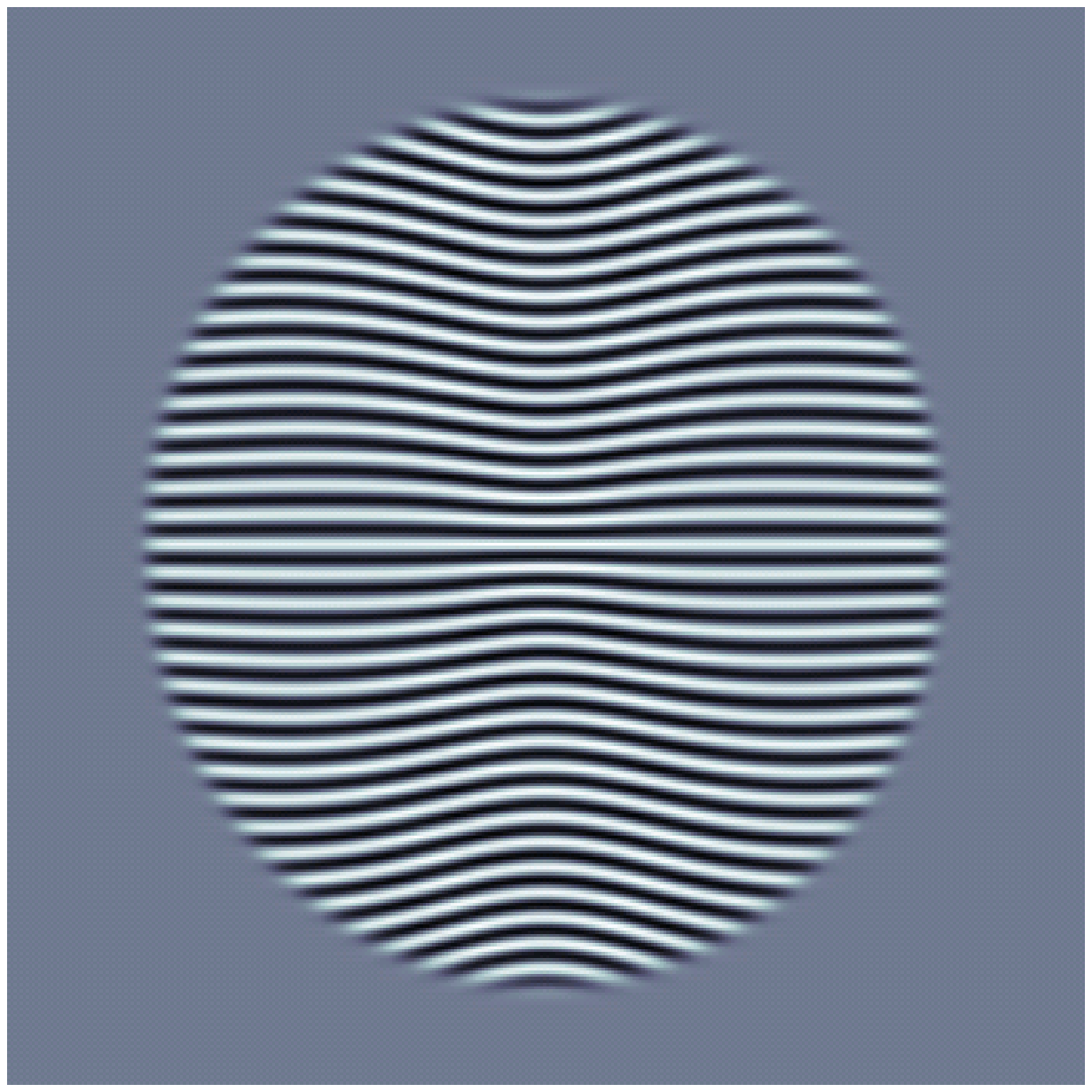}}
\subfigure[]{\includegraphics[width=0.3\linewidth]{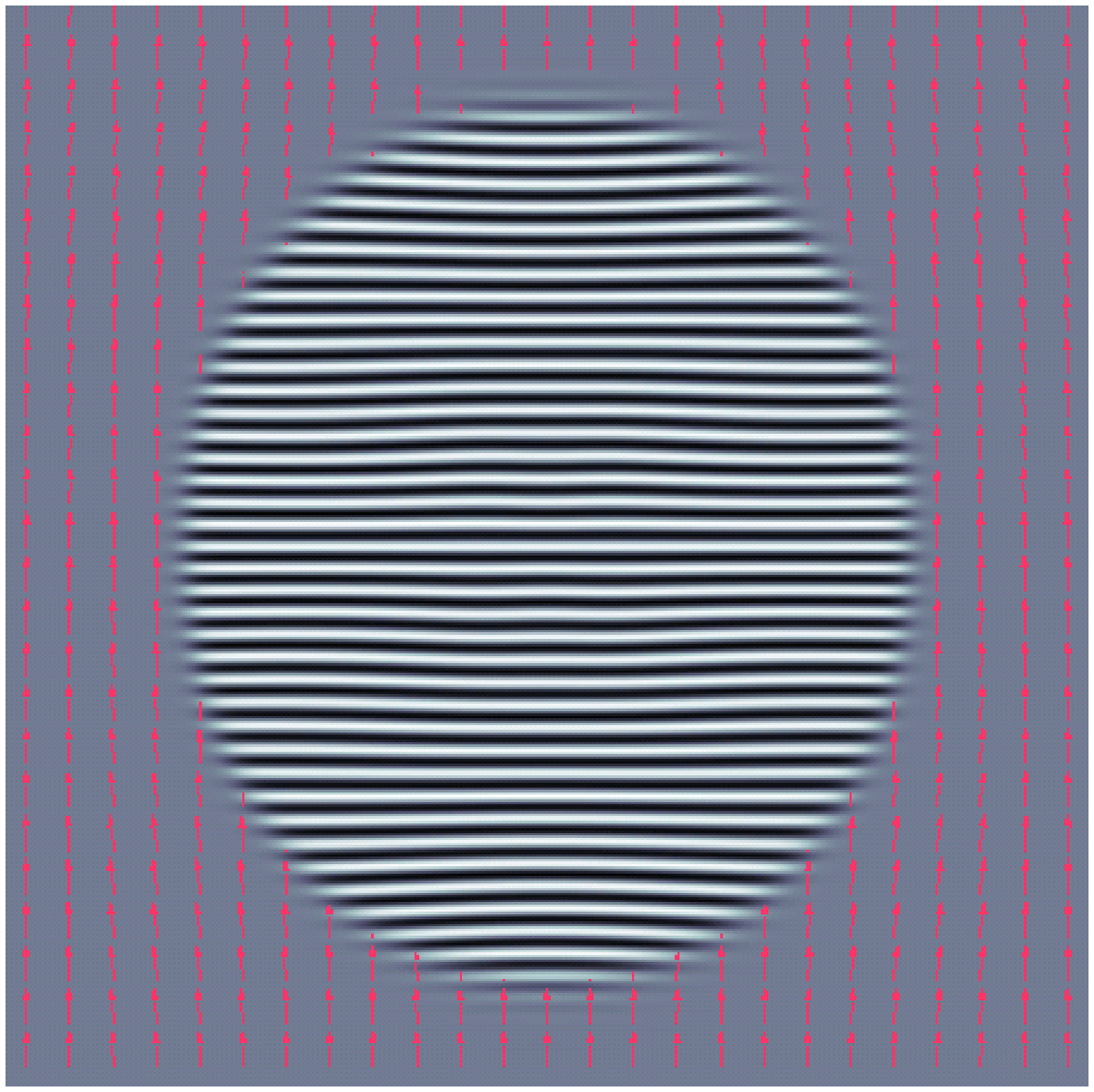}}
\subfigure[]{\includegraphics[width=0.3\linewidth]{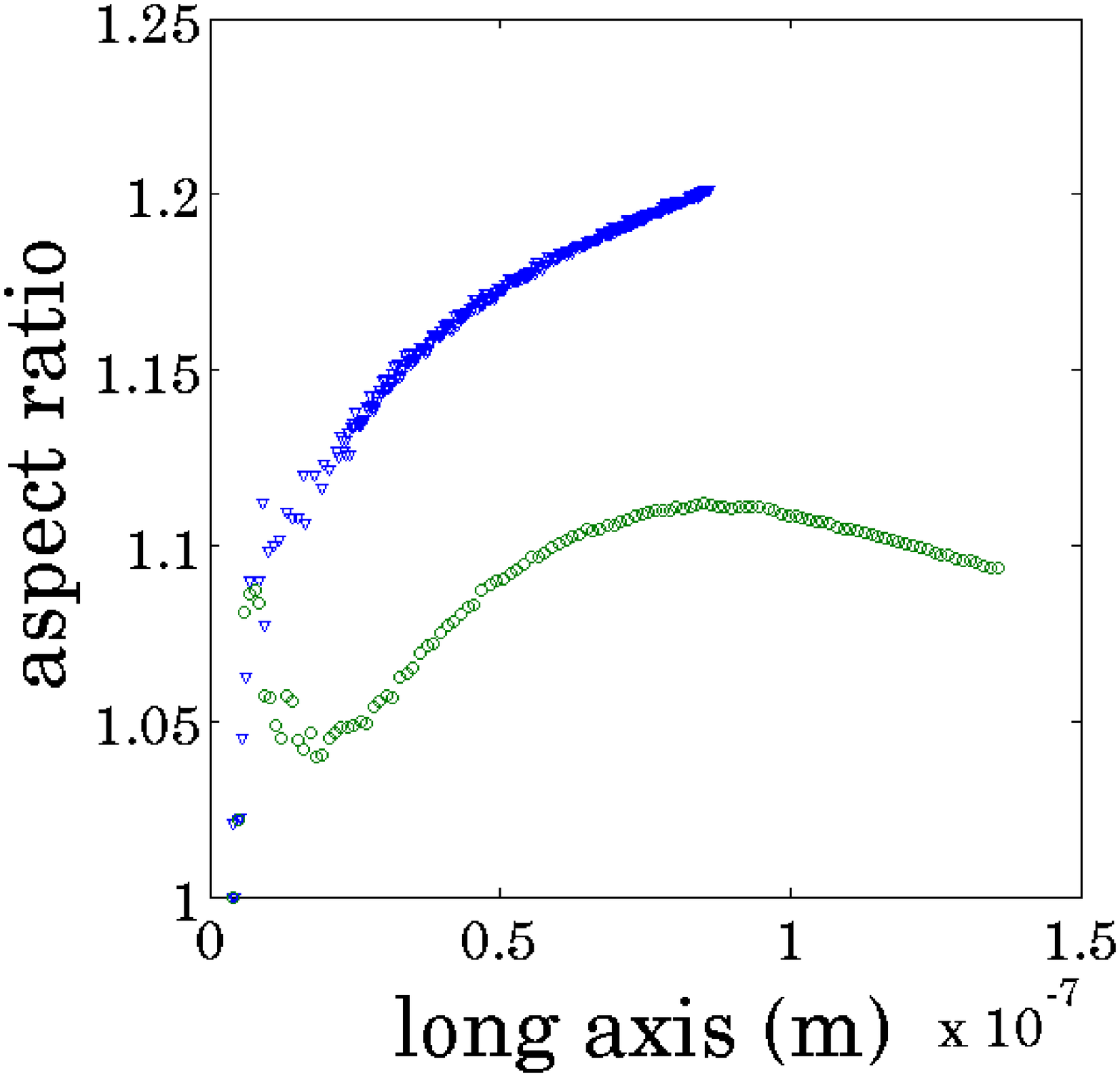}}
\caption{Two-dimensional simulation results of the growth of an initially homogeneously oriented smectic-A spherulite without a) ($\mu^* = 25$) and with meta-stable nematic pre-order b) ($\mu^* = 250$) where the surface corresponds to $Re(\Psi)$ (the smectic-A layers) and arrows denote the presence of nematic order and the director orientation (should be considered headless) c) plot of the aspect ratio versus long axis (vertical) of the simulations from a) (circles) and with pre-ordering (triangles). Other simulation parameters were used from Figure \ref{fig:plot}.}
\label{fig:sim}
\end{figure*}

In order to examine the second experimental observation, that the presence of meta-stable nematic pre-ordering induces persistent morphological changes of the liquid crystalline domain, simulation of the growth of an initially homogeneously oriented smectic-A spherulite was performed in two-dimensions. This simulation was formed in the FDC growth regime (meta-stable nematic pre-ordering) and compared to past work \cite{Abukhdeir2009} on the same system in the FC growth regime (no pre-ordering). Quasi-bulk conditions were imposed using full periodicity which results in an effective ensemble of well-oriented spherulites growing in an isotropic matrix.  Figures \ref{fig:sim}a and \ref{fig:sim}b show the spherulite morphologies at the maximum radius achievable with the current computational limitations (for the FDC simulation). The morphological growth process of the FC regime case is complex (refer to \cite{Abukhdeir2009}), but can be summarized as involving the competition between interfacial anchoring at the isotropic/smectic-A interface and bulk texturing. The resulting spherulite texture assumes a quasi-bipolar texture and exhibits an undulation instability in the bulk \cite{Abukhdeir2009}. The effect of the fully formed meta-stable nematic matrix phase (Figure \ref{fig:sim}b) has a stabilizing field effect and exhibits tactoidal growth of a neat texture. The aspect ratio versus long-axis (vertical in both cases) are shown in Figure \ref{fig:sim}c. The effects of nematic pre-ordering are shown in the current simulation to have a strong morphological effect on the smectic-A spherulite, minimizing the previously observed interfacial anchoring effects and undulation instability.  The well-oriented nematic matrix phase acts in the same way as a bulk electric/magnetic-field \cite{Das2006a}, which promotes tactoidal growth of a well-oriented smectic-A domain, consistent with the observations of Tokita et al \cite{Tokita2006}.

The aspect ratio trend for the FC regime simulation (Figure \ref{fig:sim}c) shows a decay, after an initial shape dynamic period, of the aspect ratio as the spherulite size increases. This decay is explained based upon a scaling theory derived from the studying shape and director-field transformation of nematic tactoids \cite{Prinsen2003,Prinsen2004}. A scaling estimate for the aspect ratio of the spherulite was determined to obey the following relationship to minimize total free energy \cite{Prinsen2003}:
\begin{equation} \label{eqn:bipol2}
\frac{R}{r} \approx K^{3/5}\tau^{-3/5}V^{-1/5}
\end{equation}
where $R$ is the major axis and $r$ is the minor axis, $K$ the characteristic Frank elastic constant, $\tau$ is interfacial tension, and $V$ is the volume of the spherulite.  Equation (\ref{eqn:bipol2}) predicts that the spherulite aspect ratio decreases with volume converging to $1$, in agreement with Figure \ref{fig:sim}c.

For the FDC case, the stabilizing nematic pre-ordering results in a spherulite with a neat texture and thus should obey the Wulf-construction \cite{Virga1994a} which determines surface shape by minimizing the sum of total interfacial energy and an ideal undistorted bulk contribution \cite{Virga1994a}:
\begin{equation} 
F = \int_A \gamma (\bm{r}) dA + \alpha V
\end{equation}
where $F$ is the total free energy of the spherulite, $\gamma$ is the interfacial tension (a function of position $\bm{r}$), and $\alpha$ the free energy density of the spherulite bulk. For the phenomenological parameters used in these simulations, planar anchoring (director parallel to the interface) is energetically preferred over homeotropic (director perpendicular to the interface). Figures \ref{fig:sim}b and \ref{fig:sim}c show that this approximation correctly predicts tactoidal growth, maximizing the planar-anchored interfacial area.

The simulated system is one example of the possible effects of nematic pre-ordering on the formation of a stable smectic-A domain.  Past work \cite{Wincure2007,Wincure2007a} on the growth of initially homogeneously oriented nematic spherulites implies another possible scenario.  In that work, deep quenches were found to result in the shedding of four $+\frac{1}{2}$ disclination defects in the nematic case. This is not morphologically what is proposed for the smectic-A case \cite{Abukhdeir2009}. Thus nematic pre-ordering could result in a different smectic-A domain morphology due to difference in defect shedding characteristics of the preceding nematic front, which would serve as a template. These issues will be investigated in future work.

In conclusion, it has been shown that through simulation of a high-order Landau-de Gennes type model of the isotropic/smectic-A liquid crystalline transition experimental observations of nematic pre-ordering of smectic liquid crystalline transitions \cite{Tokita2006} can be studied.  Leveraging this type of high-dimensional model has been shown to close the gap between experimental observations of this phenomena \cite{Tokita2006} and generalized theoretical studies \cite{Bechhoefer1991,Tuckerman1992}. Phase transition kinetics results presented show that nematic pre-ordering results from both thermodynamic potential and dynamic differences in phase-ordering timescales (Figure \ref{fig:plot}b), as was observed experimentally \cite{Tokita2006}.  Two-dimensional simulation results presented explain the experimentally observed effect of nematic pre-ordering on growth morphologies of smectic domains, where pre-ordering of  nematic domain stabilizes a neat texture in the growing smectic-A nucleus, promoting tactoidal growth. This work sets the basis for further simulation study in two/three-dimensions and extension of the existing model \cite{deGennes1995,Mukherjee2001} to account for the microstructure present in polymer liquid crystals. Additionally, these results, in addition to those of Tokita et al \cite{Tokita2006}, determine a possible mechanism for influencing the material properties of smectic polymer liquid crystals through the induction of meta-stable nematic pre-ordering.

\section{Acknowledgements}

This work was supported by a grant from the Natural Science and Engineering Research Council of Canada.

%%%%%%%%%%%%%%%%%%%%%%%%%%%%%%%%%%%%%%%%%%%%%%%%%%%%%%%%%%%%%%%%%%%%%
%% The appropriate \bibliography command should be placed here.
%% Notice that the class file automatically sets \bibliographystyle
%% and also names the section correctly.
%%%%%%%%%%%%%%%%%%%%%%%%%%%%%%%%%%%%%%%%%%%%%%%%%%%%%%%%%%%%%%%%%%%%%
\bibliographystyle{unsrt}
\bibliography{/home/nasser/nfs/references/references}

\begin{thebibliography}{10}

\bibitem{Imai1993}
M.~Imai, K.~Kaji, and T.~Kanaya.
\newblock {Orientation fluctuations of poly (ethylene terephthalate) during the
  induction period of crystallization}.
\newblock {\em Physical review letters}, 71(25):4162--4165, 1993.

\bibitem{Imai1994}
M.~Imai, K.~Kaji, and T.~Kanaya.
\newblock Structural formation of poly(ethylene terephthalate) during the
  induction period of crystallization. 3. evolution of density fluctuations to
  lamellar crystal.
\newblock {\em Macromolecules}, 27(24):7103--7108, 1994.

\bibitem{Matsuba1999}
Go~Matsuba, Keisuke Kaji, Koji Nishida, Toshiji Kanaya, and Masayuki. Imai.
\newblock Conformational change and orientation fluctuations of isotactic
  polystyrene prior to crystallization.
\newblock {\em Polymer Journal (Tokyo)}, 31:722--727, 1999.

\bibitem{Asano1999}
T~Asano, F.J. Balta-Calleja, and A~Flores et~al.
\newblock Crystallization of oriented amorphous poly(ethylene terephthalate) as
  revealed by x-ray diffraction and microhardness.
\newblock 40(23):6475--6484, 1999.

\bibitem{Ran2002}
S.~Ran, Z.~Wang, C.~Burger, B.~Chu, and B.S. Hsiao.
\newblock {Mesophase as the precursor for strain-induced crystallization in
  amorphous poly (ethylene terephthalate) film}.
\newblock {\em Macromolecules}, 35(27):10102--10107, 2002.

\bibitem{Li2004b}
Liangbin Li and Wim~H. de~Jeu.
\newblock Shear-induced smectic ordering in the melt of isotactic
  polypropylene.
\newblock {\em Physical Review Letters}, 92:075506/1--075506/3, 2004.

\bibitem{Soccio2007}
M.~Soccio, A.~Nogales, N.~Lotti, A.~Munari, and T.~A. Ezquerra.
\newblock Evidence of early stage precursors of polymer crystals by dielectric
  spectroscopy.
\newblock {\em Physical Review Letters}, 98(3):037801, 2007.

\bibitem{Vollrath2001}
F.~Vollrath and D.P. Knight.
\newblock Liquid crystalline spinning of spider silk.
\newblock {\em Nature}, 410:541--548, 2001.

\bibitem{Dobb1977}
MG~Dobb, DJ~Johnson, and BP~Saville.
\newblock {Supramolecular structure of a high-modulus polyaromatic fiber
  (Kevlar 49)}.
\newblock {\em Journal of Polymer Science: Polymer Physics Edition}, 15(12),
  1977.

\bibitem{Tokita2006}
Masatoshi Tokita, Kwang-Woo Kim, Sungmin Kang, and Junji. Watanabe.
\newblock Polarized light scattering and synchrotron radiation wide-angle x-ray
  diffraction studies on smectic liquid crystal formation of main-chain
  polyester.
\newblock {\em Macromolecules}, 39:2021--2023, 2006.

\bibitem{Tokita2004}
Masatoshi Tokita, Shin-ichiro Funaoka, and Junji Watanabe.
\newblock Study on smectic liquid crystal glass and isotropic liquid glass
  formed by thermotropic main-chain liquid crystal polyester.
\newblock {\em Macromolecules}, 37(26):9916--9921, 2004.

\bibitem{Dierking2003}
I.~Dierking and C.~Russell.
\newblock Universal scaling laws for the anisotropic growth of sma liquid
  crystal batonnets.
\newblock {\em Physica B (Amsterdam, Neth.)}, 325:281--286, 2003.

\bibitem{Prinsen2004}
P.~Prinsen and P.~Schoot.
\newblock Continuous director-field transformation of nematic tactoids.
\newblock {\em The European Physical Journal E: Soft Matter and Biological
  Physics}, 13(1):35--41, January 2004.

\bibitem{Bechhoefer1991}
John Bechhoefer, Hartmut L\"owen, and Laurette~S. Tuckerman.
\newblock Dynamical mechanism for the formation of metastable phases.
\newblock {\em Phys. Rev. Lett.}, 67(10):1266--1269, Sep 1991.

\bibitem{Tuckerman1992}
Laurette~S. Tuckerman and John Bechhoefer.
\newblock Dynamical mechanism for the formation of metastable phases: The case
  of two nonconserved order parameters.
\newblock {\em Phys. Rev. A}, 46(6):3178--3192, Sep 1992.

\bibitem{deGennes1995}
P.G. de~Gennes and J~Prost.
\newblock {\em The Physics of Liquid Crystals}.
\newblock Oxford University Press, New York, second edition, 1995.

\bibitem{Mukherjee2001}
P.~K. Mukherjee, H.~Pleiner, and H.~R. Brand.
\newblock Simple landau model of the smectic-a-isotropic phase transition.
\newblock {\em European Physical Journal E: Soft Matter}, 4:293--297, 2001.

\bibitem{Abukhdeir2008c}
N.M. Abukhdeir and A.D. Rey.
\newblock Simulation of spherulite growth using a comprehensive approach to
  modeling the first-order isotropic/smectic-a mesophase transition.
\newblock {\em Arxiv preprint arXiv}, arXiv:0807.4525, 2008.
\newblock Accepted for publication in Communications in Computational Physics
  12MAR2009.

\bibitem{Rey2002}
AD~Rey and M~Denn.
\newblock Dynamical phenomena in liquid-crystalline materials.
\newblock {\em Annual Review of Fluid Mechanics}, 34(1):p233 --, 2002.

\bibitem{Yan2002}
J.~Yan and A.~D. Rey.
\newblock Texture formation in carbonaceous mesophase fibers.
\newblock {\em Phys. Rev. E}, 65(3):031713, Feb 2002.

\bibitem{Rey2007}
Alejandro~D. Rey.
\newblock Capillary models for liquid crystal fibers, membranes, films, and
  drops.
\newblock {\em Soft Matter}, 3:1349 -- 1368, 2007.

\bibitem{Toledano1987}
Jean-Claude Toledano and Pierre Toledano.
\newblock {\em The Landau Theory of Phase Transitions: Application to
  Structural, Incommensurate, Magnetic, and Liquid Crystal Systems (World
  Scientific Lecture Notes in Physics)}.
\newblock World Scientific Pub Co Inc, 1987.

\bibitem{Brand2001}
Helmut~R. Brand, Prabir~K. Mukherjee, and Harald. Pleiner.
\newblock Macroscopic dynamics near the isotropic-smectic-a phase transition.
\newblock {\em Physical Review E: Statistical, Nonlinear, and Soft Matter
  Physics}, 63:061708/1--061708/6, 2001.

\bibitem{Mukherjee2002a}
Prabir~K. Mukherjee, Harald Pleiner, and Helmut~R. Brand.
\newblock Landau model of the smectic c--isotropic phase transition.
\newblock {\em The Journal of Chemical Physics}, 117(16):7788--7792, 2002.

\bibitem{Biscari2007}
P.~Biscari, M.C. Calderer, and E.M. Terentjev.
\newblock Landau de gennes theory of isotropic-nematic-smectic liquid crystal
  transitions.
\newblock {\em Phys Rev E Stat Nonlin Soft Matter Phys}, 75(5):051707, May
  2007.

\bibitem{Coles1979a}
H.~J. Coles and C.~Strazielle.
\newblock The order-disorder phase transition in liquid crystals as a function
  of molecular structure. i. the alkyl cyanobiphenyls.
\newblock {\em Molecular Crystals and Liquid Crystals}, 55:237--50, 1979.

\bibitem{Barbero2000}
G.~Barbero and Luiz~Roberto Evangelista.
\newblock {\em An Elementary Course on the Continuum Theory for Nematic Liquid
  Crystals (Series on Liquid Crystals , Vol 3)}.
\newblock World Scientific Publishing Company, 2000.

\bibitem{Wincure2006}
B.~Wincure and A.D. Rey.
\newblock Interfacial nematodynamics of heterogeneous curved isotropic-nematic
  moving fronts.
\newblock {\em The Journal of Chemical Physics}, 124(24):244902, 2006.

\bibitem{Abukhdeir2009}
N.M Abukhdeir and A.D. Rey.
\newblock Shape-dynamic growth, structure, and elasticity of homogeneously
  oriented spherulites in an isotropic/smectic-a mesophase transition.
\newblock {\em Arxiv preprint}, arXiv:0902.1544, 2009.
\newblock Accepted for publication in Liquid Crystals 09MAR2009.

\bibitem{Das2006a}
S.K. Das and A.D. Rey.
\newblock Magnetic field-induced shape transitions in multiphase polymer-liquid
  crystal blends.
\newblock {\em Macromolecular Theory and Simulations}, 15(6):469, 2006.

\bibitem{Prinsen2003}
Peter Prinsen and Paul van~der Schoot.
\newblock Shape and director-field transformation of tactoids.
\newblock {\em Phys. Rev. E}, 68(2):021701, Aug 2003.

\bibitem{Virga1994a}
E.G. Virga.
\newblock {\em Variational Theories for Liquid Crystals}.
\newblock CRC Press, 1994.

\bibitem{Wincure2007}
Benjamin Wincure and Alejandro Rey.
\newblock Growth and structure of nematic spherulites under shallow thermal
  quenches.
\newblock {\em Continuum Mechanics and Thermodynamics}, 19(1):37--58, June
  2007.

\bibitem{Wincure2007a}
B.M. Wincure and A.D. Rey.
\newblock Nanoscale analysis of defect shedding from liquid crystal interfaces.
\newblock {\em Nano Letters}, 7(6):1474--1479, 2007.

\end{thebibliography}

\end{document}